\documentstyle[12pt]{article}
\begin{document}
\renewcommand{\baselinestretch}{1.5}

\newcommand\beq{\begin{equation}}
\newcommand\eeq{\end{equation}}
\newcommand\bea{\begin{eqnarray}}
\newcommand\eea{\end{eqnarray}}
\newcommand\zb{z^*}
\newcommand\zib{z_i^*}
\newcommand\zjb{z_j^*}
\newcommand\partialzi{{\partial\over\partial z_i}}
\newcommand\partialzib{{\partial\over\partial \zib}}

\newcommand\expo{e^{-\sum_i{z_i\zib\over 4}}}

\newcommand\Sum{\sum}
\newcommand\Sumi{\sum_i^N}
\newcommand\Sumij{\sum_{i,j \ne i}^N}
\newcommand\Sumijk{\sum_{i,j\ne i,k\ne i}^N}
\newcommand\Pij{\prod_{i<j}^N}
\newcommand\prl{Phys. Rev. Lett.}
\newcommand\prb{Phys. Rev. {\bf B}}

\centerline{\bf Fractional Quantum Hall Effect States as Exact Ground States}
\vskip 1 true cm

\centerline{Ranjan Kumar Ghosh}
\centerline{\it Haldia Government College, P.O. Debhog, }
\centerline{\it Midnapore 721657, India}
\vskip .5 true cm

\centerline{Sumathi Rao \footnote{{\it e-mail
address}: sumathi@mri.ernet.in}} 
\centerline{\it Mehta Research Institute, Chhatnag Road,Jhunsi,}
\centerline{\it Allahabad 221506, India}
\vskip 2 true cm
\noindent {\bf Abstract}
\vskip 1 true cm

We construct many particle Hamiltonians for which the Laughlin and Jain
wavefunctions are exact ground states. The Hamiltonians involve
fermions in a magnetic field and with inter-particle interactions. 
For the Laughlin  wave-functions,the interactions involve 
two- and three-body correlations similar to the Chern-Simons 
interactions,  whereas for the  projected Jain wave-functions, 
$N$-body interactions (which cannot be explicitly written down 
in general) are involved. 
\vskip 1 true cm

\noindent PACS numbers:

\newpage

The phenomenon of Fractional Quantum Hall Effect (FQHE)\cite{FQHE}\ 
continues to hold interest as newer and newer types of heterojunctions
are studied at large magnetic fields\cite {NEW}. It is now well-established
that the same kind of correlations are relevant both for the integer
(I)QHE and the FQHE and that the phenomenon of FQHE can be understood
as IQHE of quasi-particles called composite fermions\cite{JAIN1}.

The original Laughlin theory was initiated by a remarkable  ansatz 
wave-function  - the Laughlin wave-function\cite{LAUGHLIN} -  
which worked very well for fractions of the form $1/(2m+1)$.
Jain recognised that the Laughlin wave-functions could
be factorised as a Jastrow factor times the Slater determinant of
one filled Landau level (LL). 
This led him to
generalise the Laughlin wave-functions to the Jain wave-functions\cite{JAIN1}
by replacing  the Slater determinant of one  filled LL by that of
$n$-filled LLs. 
This worked 
for filling fractions of the form
$n/(2mn+1)$, which included all the observed fractions. 
He interpreted these wave-functions by thinking of the
Jastrow  factors as even units of flux quanta  attached to the
electrons. Thus FQHE of electrons at fractions $n/(2mn+1)$ is equivalent
to IQHE of composite electrons (electrons with $2m$ flux units
attached) at level $n$. An apparent drawback of these wave-functions was that
they involved fermions in higher LLs, whereas at very high
magnetic fields, it is well-known that all the electrons are in the
lowest LL (LLL).  Quite remarkably, however, it was shown that despite
the presence of Slater determinants of higher LLs,
the Jain wave-functions are predominantly in the LLL
due to the multiplication by Jastrow factors. 
Moreover, by projecting\cite{PROJECT} 
these wavefunctions onto the 
LLL, wavefunctions purely in the LLL could be obtained.

There have also been attempts to understand the phenomenon of the
FQHE through field theoretic models, where the flux attachment 
takes place through a Chern-Simons (CS) field. The bosonic field theory
approach\cite{LGCS}\  explains the FQHE at the Laughlin fractions by writing
the effective field theory in terms of a bosonic field and showing
that the bosonic field Bose condenses at these fractions. The
fermionic field theory\cite{LOPEZ}\  uses the picture 
of FQHE as an 
IQHE of composite fermions {\it {\` a} la} Jain. However, in both these
field theoretic pictures, the flux attachment is through a singular
gauge transformation, and at mean field level, only the filling fractions can
be determined. The derivation of the actual Laughlin 
wavefunction  is only possible after including fluctuations beyond
the mean field picture.  
More recently, Rajaraman and Sondhi\cite{RAJARAMAN}\  
have initiated a field theoretic
approach, where the flux attachment occurs through a 
non-unitary transformation.
This picture improves upon some of the drawbacks of
the earlier field theories and here, in fact, the mean field states 
are directly the Laughlin states without recourse to fluctuations.

There have been earlier studies of models for which the Laughlin states
and Jain states were exact solutions.  Models 
\cite{HALDANE, TRUGMAN}\  with ultra-short interactions
have been shown to have the Laughlin wave-function as an exact ground state
whereas a toy model\cite {PROJECT}\  where two Landau levels 
are exactly degenerate and
all higher levels are much higher in energy has the Jain 2/5 state as
the exact ground state.

In this letter,  
we construct explicit models for electrons in an external
magnetic field and with specified interactions  between particles.
When the interactions are of a two-body and three-body type, similar
(but not identical) to the CS interaction, we show explicitly
that the Laughlin wavefunction is an exact 
ground state of the system, by
rewriting the Hamiltonian in terms of appropriate creation and annihilation
operators.

Consider the Hamiltonian given by
\bea
&H& = \Sumi {({\bf p}_i -e {\bf A})^2\over 2m} + {2\eta\over m}
    \Sumij{1\over (z_i-z_j)} (\partialzib - {i e A_z\over 2}					) \nonumber\\
   &-& {2 \eta\over m}
    \Sumij{1\over (\zib-\zjb)} (\partialzi - {i e A_z^*\over 2}
) + 
   {2 \eta^2\over m}
    \Sumijk{1\over (z_i-z_j) (z_i^* - z_k^*)}
\label{hami}
\eea
in two dimensions where ${\bf A} = {B\over 2} (y,-x)$ is the gauge potential
of the external magnetic field,  ($A_z = A_x+i A_y$ and $A_z^* = 
A_x - i A_y$),  $z_i = x_i + i y_i$ denotes the 
position of the $i^{\rm th}$ particle and $\eta$ is an odd integer. 
The second and third terms in the
Hamiltonian denote two-body interaction terms whereas the third term
denotes a  three-body interaction term. 
We now rescale the distance variable as $z\rightarrow {\tilde z} = z/l$
where $l$ is the magnetic length defined here as $l=1/\sqrt {eB}$ and 
rewrite the hamiltonian (dropping the tildes) as
\bea
 H &=& {\omega\over 2} \Sumi(-4{\partial^2\over 
       \partial z_i \partial z_i^*} + z_i\partialzi - z_i^* \partialzib
        +{z_i z_i^*\over 4})\nonumber \\  
  &+& \omega~~ 
    {\Large[} 2 \eta\Sumij{1\over (z_i-z_j)} 
    (\partialzib-{z_i\over  4}) 
   - 2\eta \Sumij{1\over (\zib-\zjb)} (\partialzi+{z^*_i\over 4})\nonumber \\ 
   &+& 2\eta^2\Sumijk{1\over (z_i-z_j) (z_i^*-z_k^*)}{\Large]}  
\label{ham1}
\eea
where $\omega =eB/m$. 

Now let us consider the action of this Hamiltonian on the  Laughlin
wave-function given by
\beq
\psi_L = \Pij (z_i - z_j)^{\eta} \expo
       \equiv f^{\eta} \expo .
\eeq
We use  the fact that $\partialzi {\rm ln} f = 
\sum_{j\ne i}{1\over (z_i-z_j)}$ 
and $\partialzib {\rm ln} {\bar f} = \sum_{j\ne i} {1\over (z_i^*-z_j^*)}$, 
(where ${\bar f}$ is the complex conjugate of $f$) 
to obtain
\beq
H \psi_L = {N\omega\over 2}  \psi_L.
\eeq
Thus, we find that the Laughlin wave-function is an
exact eigenstate of the Hamiltonian in Eq.(1) with eigenvalue
$N\omega/2$. To prove that $\psi_L$ is a  ground state,
following a general procedure\cite{RKGR}, 
we rewrite the Hamiltonian in terms of creation and annihilation 
operators\footnote{We ignore the term $\sum_{j\ne i}\delta(z_i-z_j)$ obtained 
by acting  
$\partialzib$  on $\partialzi {\rm ln}
f$ since it is irrelevant when acting on antisymmetric wave-functions.}
\bea
a_i &=& {1\over \sqrt {2}}~~(2\partialzi - 2 \eta \partialzi {\rm ln} f + 
{\zib\over 2})
\nonumber\\ {\rm and} \quad 
a_i^{\dagger} &=& {1\over \sqrt {2}}~~(-2\partialzib - 
2 \eta \partialzib {\rm ln}
                {\bar f} + {z_i\over 2})
\label{ai}
\eea
as
\beq
H = \omega \Sum_i^N a_i^{\dagger} a_i + {N\omega\over 2}.
\label{ham2}
\eeq
It is easy to check that $a_i$ for all $i$ 
annihilates the state defined by
the Laughlin wave-function, which is hence the ground state of this
Hamiltonian with the energy $N\omega/2$. (An
equation similar to $a_i \psi = 0$ has been studied earlier in the context
of anyons\cite{EZAWA}). Moreover,  $a_i$ and  $a_i^{\dagger}$ 
satisfy the canonical harmonic oscillator commutation rules even 
for non-zero $\eta$. Hence, the Hamiltonian is  exactly soluble for 
all the excited states. 

Let us now interpret the interaction terms in the Hamiltonian.
By defining $F=({\rm ln} f - {\rm ln} {\bar f})$, 
and introducing 
a  new vector (NV)  potential  ${\bf V}_i^L$ 
at the position of the
$i^{\rm th}$ particle as
\beq
{\bf V}_i^L = -i\eta{\vec \bigtriangledown}_{_i} F = 
-i\eta{\vec \bigtriangledown}_{_i}({\rm ln} f - {\rm ln}
{\bar f})
\eeq
we see that the 
interactions implied by the two- and three-body  can be interpreted 
as a gauge potential. 
The gauge potential can be equivalently 
written 
in complex notation as
\bea
V_{iz}^L &=& -2i\eta\sum_{j \ne i}^N {1\over (z_i - z_j)} 
= -2i\eta\partialzi {\rm ln} f = -2i\eta\partialzi 
({\rm ln} f - {\rm ln} {\bar f})
\nonumber \\
{\rm and}\quad  
V_{iz}^{L*} &=& -2i\eta\sum_{j \ne i}^N {1\over (\zib - \zjb)} 
 =  -2i\eta\partialzib
{\rm ln} {\bar f} =  2i\eta\partialzib ({\rm ln} f - {\rm ln} {\bar f}),
\eea
(since  $2\partialzi = {\partial\over \partial_x} - 
i {\partial \over \partial_y}$ and
$2\partialzib = {\partial\over \partial_x} + i {\partial \over 
\partial_y}$).
In terms of this gauge  potential, we find that the original Hamiltonian
in Eq.(\ref{hami})
can be rewritten as 
\beq
H = \Sumi {({\bf p}_i - e{\bf A} -  {\bf V_i})^2
\over 2m}.
\label{hamcomp}
\eeq
Note that ${\bf V}_i$ is purely real, so that there is no 
problem in interpreting it as a vector potential.
It is now clear that this Hamiltonian can be obtained 
by a singular gauge transformation from the `free'
Hamiltonian, with no interparticle interactions given by
\beq
H_0 = \sum_i^N {({\bf p}_i -e {\bf A})^2 \over 2m}.
\label{hamnon1}
\eeq
However, there are several subtleties to note here. One is that
$H_0$ has anti-analytic wave-functions as the lowest energy (LLL)
wave-functions. 
Under the transformation
\bea
\psi^* \rightarrow \psi^{'} &=& e^{\eta ({\rm ln} f-{\rm ln}{\bar f})} 
\psi^*  = 
\prod_{i<j}^N \Large( {(z_i-z_j)\over (z_i^*-z_j^*)}\Large )^{\eta} \psi^* 
\nonumber \\
{\bf V}_i^L \rightarrow {\bf V}_i^{L'} &=& 0 +  \eta
{\vec \bigtriangledown}_{_i} \Sum_{j \ne i}^N{\rm ln} { (z_i - z_j)
\over (z_i^*-z_j^*)},
\label{t1}
\eea
the `free' Hamiltonian in Eq.(\ref{hamnon1}) gets transformed to
the interacting Hamiltonian in Eq.(\ref{hamcomp}).
Secondly, without the Coulomb interaction,
the Laughlin wave-function is not the unique ground state either
of the original `free' Hamiltonian or the gauge
transformed Hamiltonian. However, we follow the general
procedure adopted in the Chern-Simons literature and
assume that the Coulomb interactions will pick the Laughlin
state to have the lowest energy. The aim of all the  
vector potential models is more
to identify new mean field theories and new order parameters.

To get a clearer understanding of this  potential,
since  it depends on the positions of other particles in the system,
it is natural to compare it with the CS gauge potential 
which also arises due to the presence of other particles.  Traditionally,
the CS potential is written in  cylindrical coordinates as
\beq
{\bf a}_i^{\rm CS} = (\eta -1)\Sum_{j \ne i}^N 
                   {{\hat z} \times ({\bf r}_i - {\bf r}_j)
                     \over |{\bf r}_i-{\bf r}_j|^2}                 
\eeq
where ${\bf r}_i$ and ${\bf r}_j$ are the position vectors of the 
particles and we are considering fermion to fermion transformation.    
However, it can also be written as 
\beq
{\bf a}_i^{\rm CS} = (\eta - 1){\vec \bigtriangledown}_{_i} 
            \Sum_{j \ne i}^N \theta_{ij}
\eeq
where $\theta_{ij}$ is the relative angle between the position vectors
of the $i^{\rm th}$ and $j^{\rm th}$ particles. 
It can be generated from  the  Hamiltonian of particles in a 
uniform magnetic field  
\beq
H_0 = \sum_i^N {({\bf p}_i + e {\bf A})^2 \over 2m}
\label{hamnon}
\eeq
(note the change in sign of $B$),
with the magnetic field adjusted so that the filling factor is $1/\eta$
by a singular gauge 
transformation, at the expense of attaching phase factors to the 
wave-function. 
Under the transformation 
\bea
\psi \rightarrow \psi^{'} &=& e^{i(\eta-1)\Sum_{j\ne i}^N \theta_{ij}} 
\psi^{} =  \prod_{i<j}^N {\Large (}{(z_i -z_j)\over |z_i-z_j|}
{\Large)}^{(\eta-1)} 
\psi \\
{\bf a}_i \rightarrow {\bf a}_i^{'} 
&=& {\bf a}_i - (\eta-1) {\vec \bigtriangledown}_{_i}
\Sum_{j\ne i}^N \theta_{ij} = 
0 - i (\eta-1) {\vec \bigtriangledown}_{_i} 
\Sum_{j\ne i}^N {\rm ln} {(z_i-z_j)\over |z_i-z_j|}.
\label{t2}  
\eea
In this case, both the original and gauge transformed Hamiltonians
have analytic wave-functions as LLL states. 

The Chern-Simons potential was  introduced in the context of 
FQHE as an exact transformation on the Hamiltonian for particles in 
an external magnetic field and interacting through a Coulomb potential.
A mean field solution of the transformed Hamiltonian is then shown
to yield the correct filling factors and the Laughlin wave-function
as a solution of the model is only obtained after including fluctuations
in the random phase approximation.

Now, let us compare our NV potential with the CS potential.
Since ${\bf V}_i^L$  can be written as
\beq
{\bf V}_i^L = -i\eta {\vec \bigtriangledown}_{_i} {\rm ln} {f\over 
{\bar f}} = 2\eta {\vec \bigtriangledown}_{_i} 
\Sum_{j \ne i}^N  ~\theta_{ij}
\eeq
we see that it is  precisely the Chern-Simons potential, except
that it interacts with a strength $2\eta$ instead of $(\eta-1)$.
So, we seem to be led to the  result that the Hamiltonian 
for which the Laughlin wave-function is an exact ground state interacts
with a Chern-Simons field with a different  charge than  the usual 
Chern-Simons
theory, where the Laughlin wave-function is not an exact solution,
but appears after performing the  mean field approximation and then 
including fluctuations.

However, this puzzle can be understood in the following way. As we 
mentioned before, 
when the interaction terms are switched off in Eq.(\ref{ai}), 
$a_i$ annihilates
anti-analytic wave-functions, which, in turn implies that the sign
of the magnetic field has been switched. This leads us to the following
scenario. We  start with the non-interacting Hamiltonian
in Eq.(\ref{hamnon})
with the magnetic field adjusted so that the filling factor is $1/\eta$.
The usual Chern-Simons procedure redistributes the total flux 
($\eta$ per particle) by
decreasing the magnetic field  to unit flux per particle and localising 
the remaining flux ($(\eta -1)$ per particle), at
the position of the particles.
Since $(\eta-1)$ is an 
even integer, 
($\eta$ is odd), fermions are converted to
composite fermions which see the appropriate magnetic field for 
integer QHE. This is the mean field approximation,
where 
the filling fractions are obtained correctly but  
the wave-function is wrong (it is, in fact, singular). 
Only when we include fluctuations
about the mean field, can we correctly  obtain the Laughlin wave-function 
as the  ground state.

But the total flux $\eta$ can also be redistributed in the 
following way. We  may have a uniform magnetic field with $-\eta$
flux per particle (equivalent to changing the sign of 
the external magnetic field and obtaining the free Hamiltonian in
Eq.(\ref{hamnon1}) instead of Eq.(\ref{hamnon})) and 
localise $2\eta$ units of flux per particle, thus 
converting them to a different kind of composite fermion. 
In other words, the Hamiltonian that we have written 
down in Eq.(\ref{hami}) is equivalent to the Hamiltonian in 
Eq.(\ref{hamnon}), which is appropriate for the FQHE, at a 
mean field level. The interesting
feature of this picture is that at the mean field level itself, 
the Laughlin wave-function emerges  
as an exact ground state. 

However, we should point out that this scenario does not lead to
a particularly useful mean field approximation. As compared to the
original Hamiltonian in Eq.(\ref{hamnon}) plus the Coulomb
interaction, the Hamiltonian in
Eq.(\ref{hami}) is not really simpler. Its only virtue is that
it can be solved exactly for the ground state.

For completeness, we also compare our vector potential with
yet another vector potential in the literature discussed by 
Rajaraman and Sondhi\cite{RAJARAMAN}. 
A quantum mechanical analogue of their field theory
involves starting  with 
the non-interacting Hamiltonian in Eq.(\ref{hamnon})
and making a non-unitary transformation on the wave-function as 
\beq
\psi \rightarrow \psi^{'} = e^{(\eta-1){\rm ln} f}\psi = 
\Pij (z_i-z_j)^{(\eta-1)}\psi
\label{wf}
\eeq
for  a fermion to fermion transformation.
(Note that the transformation function is purely analytic,
in contrast to the transformations in Eqs(\ref{t1}) and (\ref{t2}).) 
Under this transformation, the usual harmonic oscillator creation
and annihilation operators transform as 
\beq
b_i = {1\over \sqrt {2}}~~(-2\partialzib + {z_i\over 2}) \rightarrow
b'_i = b_i
\nonumber
\eeq
(since $\partialzib {\rm ln} f = 0$) and 
\beq
b_i^\dagger = {1\over \sqrt {2}}~~
(2\partialzi + {\zib\over 2}) \rightarrow 
b_i^{'\dagger} = {1\over \sqrt {2}}~~(2\partialzi + 2(\eta-1) \partialzi
{\rm ln} f + {\zib\over 2}).
\eeq
The Hamiltonian is then constructed as 
\beq
H = \Sumi  b_i^{'\dagger} b_i  + {N\omega\over 2}.
\label{hamrs}
\eeq
We identify $V^C =-i(\eta-1){\vec \bigtriangledown}_{_i}{\rm ln}f$ 
as a complex vector (CV) potential. But for the inessential difference
that the vector field introduced in Ref.\cite{RAJARAMAN} 
includes the constant external magnetic field
as well, this CV field (upto a phase redefinition) 
is the same vector field introduced by them 
in a field theoretic context. 

Thus, under the transformation on the wave-function given in 
Eq.(\ref{wf}) accompanied by the following transformation on the
complex vector field,
\beq
V^C \rightarrow V^{C'} = 0 -i (\eta-1) \Sum_{j\neq i}^N {\rm ln}(z_i-z_j),
\eeq
the non-interacting Hamiltonian in Eq.(\ref{hamnon})
gets precisely transformed to the interacting Hamiltonian in
Eq.(\ref{hamrs}). Note that the the Hamiltonian in Eq.(\ref{hamrs})
is not naively hermitean. However, by defining a new inner product
in the Hilbert space
\beq
<\psi|O|\phi> = \int \psi^* O \phi e^{-(\eta-1)({\rm ln}f + {\rm ln}
{\bar f})} d^2 x
\eeq
analogous to the field redefinitions made in Ref.\cite{RAJARAMAN},
the Hamiltonian becomes hermitean because
\beq
\int \psi^* (b_i \phi) e^{-(\eta-1)({\rm ln}f+{\rm ln}{\bar f})} d^2 x
= \int (b_i^{'\dagger} \psi)^* \phi 
e^{-(\eta-1)({\rm ln}f+{\rm ln}{\bar f})} d^2 x.
\eeq  
Hence, the Rajaraman-Sondhi Hamiltonian is hermitean although it
has been obtained by a non-unitary transformation. 

The CV  field  attaches `fat' vortices to fermions,
unlike the CS gauge field which attaches infinitesimal flux-tubes,
where only the phase is relevant.
This is closer to Jain's picture, where he attaches
`fat' vortices to electrons to turn them into composite fermions. 
The `fat' vortex factors increase the degree of the polynomial
and consequently, the area occupied by the particles in the
circular droplet. However, the magnetic field $B$ and the total
number of particles $N$ remain fixed. This means that the
density of the particles has decreased.  Hence, adding `fat' vortex
factors changes the filling fraction and is {\it not} a gauge transformation.
 
The CV  field has a representation in terms of 
the CS gauge field.
In fact, it is easy to check from the explicit definitions of the CS field and
the CV field  that
\beq
{\bf V}^C = i{\bf a}^{\rm CS} + {\hat z} \times {\bf a}^{\rm CS}.
\eeq
Thus, unlike the CS field which had only angular components, the
CV field has both radial and angular components. Note however, that
the density dependence of the CV field is solely through the CS construction.

Since the Laughlin wave-function is the exact ground state
wave-function of the Hamiltonian with particles interacting
through the ${\bf V}^L_i$ field with a strength proportional to the inverse of
the filling fraction, the bottomline question to address is why
the Coulomb interaction between particles disguise themselves
as these vector interactions. We have no insight into this problem.

What we have instead addressed are  the differences between attaching 
flux-factors $\Pi_{i<j}^N{(z_i-z_j)\over |z_i-z_j|}$  
through the CS gauge field,  the   
factor $\Pi_{i<j}^N{(z_i-z_j)\over (z_i^*-z_j^*)}$ through the 
NV field, and the vortex factor $\Pij (z_i-z_j)$ through the 
CV field.
Thus, the NV and CV Hamiltonians also (and
correspondingly the appropriate  field theories) deserve further
study to see if they  yield quantitatively correct results.

Is it possible to construct
new Hamiltonians for which the projected 
Jain wave-functions are exact ground states?
The projected Jain wave-functions are  given by
\beq
\psi_J = f_J e^{-\sum_i {z_i z_i^*\over 4}} = 
{\rm P} \Pij (z_i-z_j)^{2m} \chi_n e^{-\sum_i {z_i z_i^*\over 4}}.
\eeq
where P is the projection operator that projects the wave-function
onto the LLL. This  means that  to get the actual form of $f_J$, we need to
make the substitution $z_i^* \rightarrow 2\partialzi$ (with all $z^*$'s
to the left of the $z$'s before substitution)\cite{PROJECT}. 
Thus, $f_J$ is an  analytic
function of the $z_i$'s with terms of the form
\beq
z_1^{r_1} z_2^{r_2} \cdots z_N^{r_N} \nonumber
\eeq
with coefficients which can be explicitly determined by expanding the
Jastrow factor, the determinant and by acting with the derivatives. In
practice, this is only possible for small values of $N$. However, even
in general,
$f_J$ is a homogeneous function of its
$N$ variables, since both $\chi_n$ and the Jastrow factor are homogeneous.
Moreover, as we have already mentioned, the projection operator P
ensures that $f_J$ is analytic. For a homogeneous and analytic $f_J$,
we find that the wave-function $\psi_J$ is a ground-state
of the same Hamiltonian in Eq.(\ref{ham2}) with $a_i$ and 
$a_i^{\dagger}$ still given
by Eq.(\ref{ai}). However, now the  interaction term $\partialzi
{\rm ln} f_J = \partialzi ({\rm ln} f_J - {\rm ln}{\bar f}_J)$
can have arbitrary $N$-body interactions, since $f_J$ is an
arbitrary homogeneous and analytic function. We may still 
interpret the interaction term as a  vector  potential 
$V_{iz}^J = 2i\partialzi ({\rm ln} f_J - {\rm ln}{\bar f}_J)$, 
but this is no longer related to
the CS gauge field ${\bf a}_i^{\rm CS}$ like ${\bf V}_i^L$. More specifically,
${\bf V}_i^J$ is not constrained by the density of particles alone, but
by more complicated correlations. This explains why these states
have not been amenable to the usual CS - type field theoretic treatments,
although they have the virtue of analyticity. Much further work 
is needed to understand 
the nature of this kind of NV potential. 

In conclusion, in this letter, we have shown that the Laughlin
wave-function is an exact ground state of a Hamiltonian with many 
particles in an external magnetic field and with inter-particle
interactions that can be described by a NV field related to the density
of particles. We have also compared our vector potential
with the Chern-Simons and Rajaraman-Sondhi vector potentials  
and discussed the merits and demerits of the
different approaches.

\section*{Acknowledgments}

One of us (S.R.) would like to thank D. M. Gaitonde and R. Rajaraman
for useful discussions. RKG would like to thank  the Mehta Research Institute
for hospitality during the initiation of this work.

\end{document}